\documentclass[a4paper]{ifacconf}%a4paper letterpaper

\usepackage{graphicx}      % include this line if your document contains figures
\usepackage{natbib}        % required for bibliography
\usepackage{amsmath}
\usepackage{amssymb}
\usepackage{multirow}
\usepackage{enumerate}
\usepackage{tikz}
\usepackage{algorithm}
\usepackage[]{algpseudocode}
\newcommand{\pa}{{{\boldsymbol \theta}}}
\newcommand{\est}{{{\hat{\pa}}}}
\newcommand{\gt}{{ {\pa_{\rm t}}}} %\scriptstyle
\newcommand{\parpa}{{\partial {\pa}}}%\scriptstyle
\newcommand{\0}{\boldsymbol 0}%\scriptstyle
\newcommand{\x}{\boldsymbol x}%\scriptstyle
\newcommand{\z}{\boldsymbol z}%\scriptstyle
\newcommand{\f}{\boldsymbol f}%\scriptstyle
\newcommand{\U}{\boldsymbol U}%\scriptstyle
\newcommand{\V}{\boldsymbol V}%\scriptstyle
\newcommand{\M}{\boldsymbol M}%\scriptstyle
\newcommand{\N}{\boldsymbol N}%\scriptstyle
\newcommand{\idx}{{\rm idx}}%\scriptstyle
\newcommand{\st}{{\rm s.t.}}%\scriptstyle

\usepackage{float}
\newfloat{algorithm}{t}{lop}

\newcommand\copyrighttext{%
  \footnotesize \centering
  \noindent 
\textcopyright\hspace{-0.1cm} 2020 the authors. %Marcel Menner and Melanie N. Zeilinger.   %IFAC
\textit{This work has been accepted to IFAC for publication under a Creative Commons Licence CC-BY-NC-ND}.
   }
\newcommand\copyrightnotice{%
\begin{tikzpicture}[remember picture,overlay]
\node[anchor=south,yshift=10pt] at (9,-15.6) {{\parbox{\dimexpr\textwidth-\fboxsep-\fboxrule\relax}{\copyrighttext}}};
\end{tikzpicture}%
}
\begin{document}

%\input{arxiv_page}
%\newpage
\begin{frontmatter}

\title{Maximum Likelihood Methods for Inverse Learning of Optimal Controllers\thanksref{footnoteinfo}} 

\thanks[footnoteinfo]{This work was supported by the Swiss National Science Foundation under grant no. PP00P2{\_}157601 / 1.}

\author[First]{Marcel Menner} \textbf{and} 
\author[First]{Melanie N. Zeilinger}

\address[First]{Institute for Dynamic Systems and Control, ETH Zurich, \\
Zurich, Switzerland 
(e-mail: \{mmenner,mzeilinger\}@ethz.ch)}

\begin{abstract}                
This paper presents a framework for inverse learning of objective functions for constrained optimal control problems, which is based on the Karush-Kuhn-Tucker (KKT) conditions. We discuss three variants corresponding to different model assumptions and computational complexities. The first method uses a convex relaxation of the KKT conditions and serves as the benchmark. The main contribution of this paper is the proposition of two learning methods that combine the KKT conditions with maximum likelihood estimation. The key benefit of this combination is the systematic treatment of constraints for learning from noisy data with a branch-and-bound algorithm using likelihood arguments. This paper discusses theoretic properties of the learning methods and presents simulation results that highlight the advantages of using the maximum likelihood formulation for learning objective functions.
\end{abstract}

\begin{keyword}
Learning for control,
data-based control,
constrained control.
\end{keyword}

\end{frontmatter}
%===============================================================================

\copyrightnotice
\vspace{-0.5cm}

\section{Introduction} 
Objective functions used for control design do not necessarily correspond to the actual performance specifications for a dynamical system, which may comprise complex or sparse targets.
Instead, they are often chosen to facilitate gradient-based numerical optimization, which, in turn, makes their design not very intuitive and their calibration can require a tedious manual engineering effort to meet the performance specifications. 
Inverse learning concepts such as inverse optimal control offer an attractive design paradigm for learning objective functions from data to avoid their manual tuning.
In this context, the data can originate, e.g., from a human actor, who demonstrates how to optimally operate the dynamical system being considered. 
Learning from demonstrations, however, necessarily implies that the data are subject to noise and other sources of sub-optimalities, which have to be taken into account.

In this paper, we present and contrast three variants of an inverse optimal control approach that leverages the Karush-Kuhn-Tucker optimality conditions, cf. \cite{Karush1939,Kuhn1951}, to learn objective functions of optimal controllers, e.g., for linear quadratic or model predictive control, from noisy data. 
The first method is based on a convex relaxation of the KKT conditions to allow for noisy data, which is similar to the formulation in \cite{Englert2017,menner2018predictive} and included in this paper as the benchmark.
The main contribution of this paper is the proposition of two inverse optimal control methods that combine the KKT conditions with a maximum likelihood estimation algorithm, which offer the key benefit of systematically dealing with state and input constraints in the presence of noisy data. 
The underlying assumption is that the data are samples from a distribution (rather than expecting deterministic, optimal data).
Maximum likelihood estimation is enabled by an algorithm that uses branch-and-bound-type ideas based on the likelihoods of active constraints. 
The second contribution is a theoretical and simulative analysis of the properties of the three methods. 
In theory, we analyze the learning results of the inverse optimal control methods for unconstrained, linear dynamical systems and a quadratic cost function.
In simulation, we present learning results for both constrained, linear and nonlinear systems.

\subsubsection*{Related work}
Inverse optimal control methods typically model data as deterministic and resulting from an optimal control problem [\cite{Hewing2020}], 
whereas we explicitly consider the data as stochastic.
In \cite{Kalman1964}, \cite{Menner2018ecc}, and \cite{mombaur2010human}, inverse optimal control methods for linear, unconstrained systems are presented.
\cite{Englert2017} and \cite{menner2018predictive} use a formulation similar to the first method presented in this paper, which is based on the relaxation of the KKT conditions.
\cite{Chou2018constraints,chou2020} address a related problem by learning constraints. 
The method in \cite{menner2019cdc} considers a non-deterministic model, but does not consider constraints in the learning procedure.
The closest to the proposed likelihood estimation methods is \cite{esfahani2018data}, 
where the main difference lies in the proposed formulation using likelihood arguments offering the key advantage of dealing with constraints using a branch-and-bound algorithm.

Inverse reinforcement learning methods typically model data by means of a Markov decision process, cf. \cite{Ziebart2008,levine2012continuous,Finn2016}.
As a result, these methods can deal with noise by construction, but constraints are typically not considered.
Compared to inverse reinforcement learning methods, we base our algorithm on the KKT conditions in order to explicitly consider constraints and noisy data.

\section{Problem Statement}
We consider discrete-time dynamical systems of the form
\begin{align}
\label{ifac:eq:sys}
\boldsymbol x(k+1) = \boldsymbol f(\boldsymbol x(k),\boldsymbol u(k)),
\end{align}
where $\boldsymbol x(k)\in \mathbb{R}^n$ is the state at time $k$, $\boldsymbol u(k)\in \mathbb{R}^m$ is the input, and $\boldsymbol f$ is, in general, a nonlinear function.

\subsubsection{Control Model}
We consider optimal controllers of the form
\begin{equation}
\label{ifac:eq:forward}
\begin{aligned}
\boldsymbol v_k^\star 
=
\arg \hspace{-0.25cm} \min_{\boldsymbol v_k,\boldsymbol z_k\ \forall k}\ &  
\pa^T
\boldsymbol \phi
\left(
\boldsymbol v_0,...\boldsymbol v_{N},
\boldsymbol z_0,...\boldsymbol z_{N+1}
\right)
\\
{\rm s.t.}\ 
& 
\boldsymbol z_{k+1} = \boldsymbol f(\boldsymbol z_k, \boldsymbol v_k) \quad \forall k=0,...N
\\
&
\boldsymbol g(\boldsymbol z_k, \boldsymbol v_k)\leq \boldsymbol 0 \quad\quad \forall k=0,...N
\\
& \z_0 = \z(0),
\end{aligned}
\end{equation}
where $\boldsymbol v_k^\star$ are optimal inputs at time $k$, $\boldsymbol z_k$ are the predicted states given inputs $\boldsymbol v_k$, and the initial condition is $\z(0)$.
The minimizers of \eqref{ifac:eq:forward}, i.e., the nominal states $\z_k^\star$ and inputs $\boldsymbol v_k^\star$, express a motion plan and do not necessarily coincide with the measured states $\x(k)$ and inputs $\boldsymbol u(k)$.
The objective function is defined by $\boldsymbol \phi$, which is weighted by the parameters $\boldsymbol \theta$.
The function $\boldsymbol g$ defines constraints and $N$ is the prediction horizon.
We assume that $\boldsymbol \phi$, $\boldsymbol f$, and $\boldsymbol g$ are known and continuously differentiable.

\subsubsection*{Assumption on the data}
In expectation, the data are assumed to be the solution to \eqref{ifac:eq:forward}, i.e., the demonstration, e.g. from a human agent, is modeled as an optimal controller.
Due to noise and other sources of sub-optimalities, we assume a probability distribution for the data:
\begin{subequations}
\vspace{-.15cm}
\label{ifac:eq:assumptionsData}
\begin{enumerate}[i)]
\item We model the initial condition, denoted $\x(0)$, as uncertain and assume 
\begin{equation}
\label{ifac:eq:initDist}
\x(0) \sim \mathcal N(\z(0),\boldsymbol  \Sigma_0),
\end{equation}
i.e., $\x(0)$ is Gaussian distributed with mean $\z(0)$ and covariance $\boldsymbol \Sigma_0$.
\item We model the observed inputs, denoted $\boldsymbol u(k)$, as suboptimal and assume 
\begin{equation}
\label{ifac:eq:control}
\boldsymbol u(k)
\sim
\mathcal{N}
(\boldsymbol v_k^\star, \boldsymbol \Sigma^u_k)\quad \forall k=0,...,N,
\end{equation}
where $\boldsymbol v_k^\star$ are the minimizers of \eqref{ifac:eq:forward}.
\end{enumerate}
\end{subequations}
In the context of learning from data generated by a human agent, eq. \eqref{ifac:eq:initDist} and eq. \eqref{ifac:eq:control} model that a human agent may be uncertain about the true initial state $\x(0)$ and may not execute the intended motion plan optimally.

\subsubsection{Objective}
In this paper, we learn the parameters $\pa$ of the optimal controller in \eqref{ifac:eq:forward} from data represented in the form of state $\x(k)$ and input $\boldsymbol u(k)$ measurements satisfying \eqref{ifac:eq:sys} generated, e.g., by a human actor modeled as in \eqref{ifac:eq:assumptionsData}. 

\subsection*{Notation \& Preliminaries}
\vspace{-.25cm}
In order to ease exposition, we vectorize sequences
\begin{align*}
\boldsymbol U
=
\begin{bmatrix}
\boldsymbol u(0)
\\
\boldsymbol u(1)
\\
\vdots
\\
\boldsymbol u({N})
\end{bmatrix}
,\
\boldsymbol Z
=
\begin{bmatrix}
\boldsymbol z_0
\\
\boldsymbol z_1
\\
\vdots
\\
\boldsymbol z_{N+1}
\end{bmatrix}
,\
\boldsymbol V
=
\begin{bmatrix}
\boldsymbol v_0
\\
\boldsymbol v_1
\\
\vdots
\\
\boldsymbol v_{N}
\end{bmatrix}
\end{align*}
and use the function $\boldsymbol F$ relating the vectorized sequences as in \eqref{ifac:eq:sys}:
$ 
\boldsymbol Z
=
\boldsymbol F
(\boldsymbol V,\boldsymbol z_0).
$ 
We use 
$\boldsymbol \phi
\left(
\boldsymbol v_0,...\boldsymbol v_{N},
\boldsymbol z_0,...\boldsymbol z_{N+1}
\right) = 
\boldsymbol \phi
\left(
\boldsymbol V,
\boldsymbol Z
\right)$, as well as 
$\boldsymbol g
\left(
\boldsymbol v_0,...\boldsymbol v_{N},
\boldsymbol z_0,...\boldsymbol z_{N+1}
\right) = 
\boldsymbol g
\left(
\boldsymbol V,
\boldsymbol Z
\right)$
equivalently.
Further, 
$\boldsymbol U
\sim
\mathcal{N}
(\boldsymbol V, \boldsymbol \Sigma)$
implies
$\boldsymbol u(k) 
\sim
\mathcal{N}
(\boldsymbol v_k^\star, \boldsymbol \Sigma^u_k)$ for all $k=0,...,N$, i.e., $\boldsymbol \Sigma\in \mathbb{R}^{mN\times mN}$ is block-diagonal with blocks $\boldsymbol \Sigma^u_k$
and we define $\|\boldsymbol x\|_{\boldsymbol X}=\boldsymbol x^T \boldsymbol X \boldsymbol x$.

Let $\idx, \neg \idx\in \{0,1\}^s$ with $\idx + \neg \idx = \{1\}^s$.
For a vector $\boldsymbol \lambda\in \mathbb{R}^s$, we define $\boldsymbol \lambda_{\rm idx}$ selecting all elements $\lambda_i$ for which $\idx_i=1$ ($\boldsymbol \lambda_{\neg \idx}$ selecting $\lambda_i$ for which $\idx_i=0$). 
$\boldsymbol \delta^{i}$ is a unit vector with $\delta^{i}_j=1$ if $i=j$ and $\delta^{i}_j=0$ if $i\neq j$.

Consider the optimization problem
\begin{equation}
\label{ifac:eq:prelim}
\begin{aligned}
\V^\star=\arg\min_{\V}\ & \pa^T \boldsymbol \phi(\V,\x)
\\
{\rm s.t.}\ & \boldsymbol g(\V,\x)\leq \0.
\end{aligned}
\end{equation}
The KKT conditions of \eqref{ifac:eq:prelim} are given by
\begin{align}
\label{ifac:eq:KKT}
{\rm KKT}_{\boldsymbol \theta}(\V^\star,\x)
= 
\begin{cases}
\nabla_{\V} \mathcal L_{\pa}(\V,\x) |_{\V=\V^\star}=\0
\\
\boldsymbol \lambda^T  \boldsymbol g(\V,\boldsymbol F(\boldsymbol V,\x)) = 0
\\
\boldsymbol  g(\boldsymbol V,\boldsymbol F(\boldsymbol V,\x))\leq \boldsymbol 0
 \\
\boldsymbol \lambda\geq \boldsymbol 0
\end{cases}
\end{align}
with the dual variables $\boldsymbol \lambda$ and the Lagrangian 
$$\mathcal L_{\boldsymbol \theta}({\boldsymbol V},\x) =\boldsymbol \theta^T \boldsymbol \phi({\boldsymbol V},\boldsymbol F(\boldsymbol V,\x)) + \boldsymbol \lambda^T \boldsymbol g(\V,\boldsymbol F(\boldsymbol V,\x)).$$
The KKT conditions are necessary for constrained optimization (first-order derivative tests), i.e., any $\V^\star$ locally minimizing \eqref{ifac:eq:prelim} satisfies \eqref{ifac:eq:KKT}. 
For more details, the reader is referred, e.g., to \cite{boyd2004convex}.

\textit{Proposition~1}.\quad 
Consider
\begin{align*} 
f^1=\max_{\x\in \mathbb R^{n}}\ & f(\x)
& 
f^2 = \max_{\x\in \mathbb R^{n}}\ & f(\x)
\\
\st\ & g_1(\x)\leq 0
&
\st\ & g_1(\x)\leq 0 ,g_2(\x)\leq 0
\end{align*}
and let $\{\x | g_1(\x)\leq 0\}$ be a non-empty set and $f(\x)$ be bounded.
Then, $f^1\geq f^2$.

\section{Inverse Learning Methods}
\label{sec:models}
This section presents three methods for inverse learning of the objective function, which utilize the KKT conditions in \eqref{ifac:eq:KKT} as follows:
Suppose $\V^\star$ is the result of \eqref{ifac:eq:forward} for some \underline true parameters $\pa=\gt$ and initial condition $\z(0)$.
Then, ${\rm KKT}_{\pa}(\V^\star,\z(0))$ hold for $\pa=\gt$.
Hence, the KKT conditions can be used to learn $\gt$ given $\V^\star$.
Method~1 is based on a relaxation of the KKT conditions in \eqref{ifac:eq:KKT} to allow for noisy data.
Methods~2 and 3 are based on maximum likelihood estimation and use the distribution in \eqref{ifac:eq:assumptionsData}.
The three methods vary in computational complexity and model assumptions in the form of approximations.
 
\subsubsection{Method 1}
This method uses a relaxation of the KKT conditions to directly relate the data $\U,\x(0)$ with $\pa$:
\begin{subequations}
\begin{equation}
\label{ifac:eq:naive}
\begin{aligned}
\est=\arg \min_{\pa,\boldsymbol \lambda}\ & 
\left\|
\nabla_{\V} \mathcal L_{\pa}(\V,\x(0)) |_{\V=\U}
\right\|_I 
\\
{\rm s.t.}\ &
\boldsymbol \lambda_{}\geq \0
\\
 &
\boldsymbol \lambda_{\neg \idx} = \0,
\end{aligned}
\end{equation}
where $\neg \idx$ indexes inactive constraints, $g_i(\U,\x(0))$, with  
\begin{align*}
\neg \idx_i 
=
\begin{cases}
1 & {\rm if}\ g_i(\U,\x(0))<0
\\
0 & {\rm else}.
\end{cases}
\end{align*}
The main advantage is that \eqref{ifac:eq:naive} is a convex optimization problem.
Compared to \eqref{ifac:eq:KKT}, $\nabla_{\V} \mathcal L_{\pa}(\V,\x(0)) |_{\V=\U}\neq \0$ as well as $\boldsymbol g(\U,\boldsymbol F(\boldsymbol U,\x)) \neq \0$ due to noisy data.
Therefore, we minimize $\nabla_{\V} \mathcal L_{\pa}(\V,\x(0)) |_{\V=\U}$, where $\boldsymbol \lambda^T  \boldsymbol g(\V,\boldsymbol F(\boldsymbol V,\x)) = 0$ with $\boldsymbol \lambda\geq \0$ is relaxed to $\boldsymbol \lambda\geq \0$ and $\boldsymbol \lambda_{\neg \idx}= \0$.

\subsubsection{Method 2} 
This method is based on maximum likelihood estimation and uses the expected value of the initial condition in \eqref{ifac:eq:initDist} with $\x(0) =\z(0)$.
Using the distribution of the control inputs in \eqref{ifac:eq:control}, the parameters $\pa$ are estimated to maximize the probability of observing $\U$:
\begin{equation}
\label{ifac:eq:inverse}
\begin{aligned}
\est=\arg&\max_{\boldsymbol \theta}\  p(\V | \U ,\boldsymbol  \Sigma)%p(\mu_0 | x_0,\Sigma_0)
\\
&\hspace{0.1cm} 
\begin{aligned}
{\rm s.t.}\  
\boldsymbol V = 
\arg\min_{\tilde{\V}}&\
  \boldsymbol \theta^T 
  \boldsymbol  \phi
  (\tilde{\V},
  \boldsymbol F(\tilde{\V},\x(0))
)
\\
 {\rm s.t.}\ & 
 \boldsymbol g(\tilde{\V},\boldsymbol F(\tilde{\V},\x(0)))\leq \0. 
\end{aligned}
\end{aligned}
\end{equation}

\subsubsection{Method 3} 
This method considers the uncertainty about the initial condition explicitly.
The method additionally optimizes over the initial condition with $\x(0) \sim \mathcal N(\z(0),\boldsymbol  \Sigma_0)$ and the model for the inverse learning problem is given by
\begin{equation}
\label{ifac:eq:inverse2}
\begin{aligned}
\est=\arg&\max_{\boldsymbol \theta,\boldsymbol z(0)}\  p(\boldsymbol V | \boldsymbol U,\boldsymbol  \Sigma)p(\boldsymbol z(0) | \boldsymbol x(0),\boldsymbol \Sigma_0)
\\
&\hspace{0.1cm} 
\begin{aligned}
{\rm s.t.}\  
\boldsymbol V  
= 
\arg\min_{\tilde{\V}}&\
  \pa^T 
  \boldsymbol  \phi
  (\tilde{\boldsymbol V},
  \boldsymbol F(\tilde{\boldsymbol V},{\boldsymbol z}(0))
)
\\
 {\rm s.t.}\ & 
 \boldsymbol g(\tilde{\boldsymbol V},\boldsymbol F(\tilde{\V},{\z}(0)))\leq \0.
\end{aligned}
\end{aligned}
\end{equation}
\end{subequations}

Both maximum likelihood estimation methods \eqref{ifac:eq:inverse} and \eqref{ifac:eq:inverse2} yield bi-level optimization problems that are solved as described in Section~\ref{sec:alg}.

\section{Algorithm for Maximum Likelihood Estimation}
\label{sec:alg}

In the following, we outline the algorithm for learning $\pa$ using Method~2.
The algorithm for Method~3 follows analogously.
We first replace the likelihood $p(\V | \U ,\boldsymbol  \Sigma)$ by its logarithmic likelihood $\log p(\V | \U ,\boldsymbol  \Sigma)$ (log-likelihood) and the lower level optimization problem in \eqref{ifac:eq:inverse} by its KKT conditions in \eqref{ifac:eq:KKT}.
This way, the bi-level optimization problem is replaced by a combinatorial problem due to the complementary slackness condition $\boldsymbol \lambda^T  \boldsymbol g(\V,\boldsymbol F(\boldsymbol V,\x)) = 0$: 
\begin{equation}
\label{ifac:eq:final_calc}
\begin{aligned}
p = 
\max_{
\pa,\V,\boldsymbol \lambda, \idx
}\ %
& 
\log p\left(\V \middle| \U, \boldsymbol \Sigma\right) 
\\
\st\ & 
{\rm KKT}_{\pa,\idx}(\V,\x(0))
\\
& \idx \in \{0,1\}^s 
\end{aligned}
\end{equation}
with $\idx$ selecting which of the $s$ constraints are active, i.e.
\begin{align*}
{\rm KKT}_{\pa,\rm idx}(\V,\x)
= 
\begin{cases}
\nabla_{\tilde \V} \mathcal L_{\pa}(\tilde \V,\x) |_{\tilde \V=\V}=\0
\\
\boldsymbol  g(\V,\boldsymbol F(\V,\x))\leq \0
\\
\boldsymbol  g(\V,\boldsymbol F(\V,\x))_{{\rm idx}}= 0
 \\
\boldsymbol \lambda_{}\geq \0
 \\
\boldsymbol \lambda_{\neg {\rm idx}}= \0.
\end{cases}
\end{align*}
Solving the combinatorial optimization problem in \eqref{ifac:eq:final_calc} directly is computationally intensive.
However, using likelihood arguments with branch-and-bound-type ideas, \eqref{ifac:eq:final_calc} becomes practically feasible as outlined in the following.

Algorithm~\ref{ifac:alg:ML} summarizes the procedure to solve \eqref{ifac:eq:final_calc}, which is based on systematically enumerating candidate solutions and is conceptually similar to active-set methods, cf., \cite{murty1988linear}.
The algorithm aims at reducing the number of times \eqref{ifac:eq:final_calc} has to be solved for a fixed combination of active constraints, denoted $\idx^j\in \{0,1\}^s$, where we use $j$ to index the specific combination of active constraints.
First (Line~\ref{ifac:alg:p0} in  Alg.~\ref{ifac:alg:ML}), we solve 
\begin{equation}
\label{ifac:eq:initial_calc}
\begin{aligned}
p^0 = 
\max_{\pa,\V}\ 
& 
\log p\left(\V \middle| \U, \boldsymbol \Sigma\right) 
\\
{\rm s.t.}\ & 
{\rm KKT}_{\pa,{\rm idx}^0=\{0\}^s}(\V,\x(0)),
\end{aligned}
\end{equation}
where $p^0$ is the log-likelihood of observing $\U$ and no active constraints (${\rm idx}^0=\{0\}^s$).
Next (Line~\ref{ifac:alg:barp}), we compute an upper bound on the log-likelihood of constraint $i$'s activeness given the data $\U$ as 
\begin{equation}
\label{ifac:eq:constraint_likelihood}
\begin{aligned}
\bar
p^i
=
\max_{\boldsymbol V%,\boldsymbol z_0
}\ &
\log
p(\V | \U, \boldsymbol \Sigma)  
\\
{\rm s.t.}\ & 
\boldsymbol g(\V,\boldsymbol F(\V,\boldsymbol x(0))) \leq \0
\\& 
\boldsymbol g(\V,\boldsymbol F(\V,\boldsymbol x(0)))_{{\boldsymbol \delta^{i}}}=\0.
\end{aligned}
\end{equation}
From Proposition 1, constraint $i$ is not likely to be active 
if $\bar p{}^i\leq p^0$ and consequently, constraint $i$ does not need to be enumerated, which is the first key component of the algorithm's efficiency as the number of possible constraint combinations, denoted $c$,  can be reduced significantly.

Next (Line~\ref{ifac:alg:candset}), we compute upper bounds for the log-likelihood $p$ in \eqref{ifac:eq:final_calc} for the fixed combinations of the active constraints $\idx^j$ (excluding the discarded constraints):
\begin{equation}
\label{ifac:eq:comp_cand}
\begin{aligned} 
\tilde p{}^j
=
\max_{\boldsymbol V%,\boldsymbol z_0
}\ & \log p(\V | \U, \boldsymbol \Sigma) 
\\
{\rm s.t.}\ & 
\boldsymbol g(\V,\boldsymbol F(\V,\boldsymbol x(0))) \leq \0
\\& 
\boldsymbol g(\V,\boldsymbol F(\V,\boldsymbol x(0)))_{\idx^j}=\0
\end{aligned}
\end{equation}
for all $j=1,...c$.
As a result, we obtain $c$ possible candidate constraint combinations as well as their upper bounds:
\begin{align}
\label{ifac:eq:sorted}
\mathcal D
=
\left\{
\left\{
{\rm idx}{}^1,
\tilde p{}^1
\right\},
\left\{
{\rm idx}{}^2,
\tilde p{}^2
\right\},...,
\left\{
{\rm idx}^c,
\tilde p{}^c
\right\}
\right\}.
\end{align}
For ease of exposition, $\mathcal D$ is ordered so that $\tilde p{}^j\geq \tilde p{}^{j+1}$.
The log-likelihoods $\tilde p{}^j$ are the second key component for the algorithm's efficiency and are used as the stopping criteria, i.e., \eqref{ifac:eq:final_calc} is solved for $\idx^j$ starting with $j=1$ until $\tilde p{}^j \leq \max\{p^0,...,p^{j-1} \}$ (Line~\ref{ifac:alg:step-i1}--\ref{ifac:alg:step-i2}).

\begin{algorithm}[h]
\caption{  
Overall algorithm
} 
\label{ifac:alg:ML}
\begin{algorithmic}[1]
\State $\hat \pa,  p^0 \leftarrow$ Solve \eqref{ifac:eq:initial_calc} with ${\rm idx}=\{0\}^s$ 
\label{ifac:alg:p0}
\State For each $i=1,...,s$, compute $\bar p{}^i$ using \eqref{ifac:eq:constraint_likelihood} and discard constraint $i$ if $\bar p{}^i\leq p^0$
\label{ifac:alg:barp}
\State Compute $\mathcal D$ in \eqref{ifac:eq:sorted} using \eqref{ifac:eq:comp_cand}
\label{ifac:alg:candset}
\State $j=1,\hat p=p^0$
\State \textbf{while} $\tilde p{}^j > \max\{p^0,...,p^{j-1} \}=\hat p$ \Comment{end if less likely}
\label{ifac:alg:step-i1}
\State \quad $\pa^j, p^{j} \leftarrow$ Solve \eqref{ifac:eq:final_calc} with  fixed active constraints $\idx^j$
\State \quad \textbf{if} $p^j\geq \hat p$
\State \quad  \quad $\hat \pa\leftarrow \pa^i$, $\hat p \leftarrow p^i$
\State \quad $j\leftarrow j+1$
\label{ifac:alg:step-i2}
\end{algorithmic}
\end{algorithm}

\begin{rem}
We implemented a projected gradient method that uses backtracking line search [\cite{armijo1966minimization}] to solve both \eqref{ifac:eq:naive} for Method~1 and \eqref{ifac:eq:final_calc} with the fixed active constraints $\idx^j$ for Method~2 and Method~3.
Section~\ref{ifac:sec:results} details the computation times of the three inverse learning methods, which show that the proposed algorithm is computationally feasible.
\end{rem}

\begin{figure}[h]
\begin{center}
\includegraphics[trim={0.5cm 0.3cm 0.5cm 0.8cm},clip,width=0.99\columnwidth]{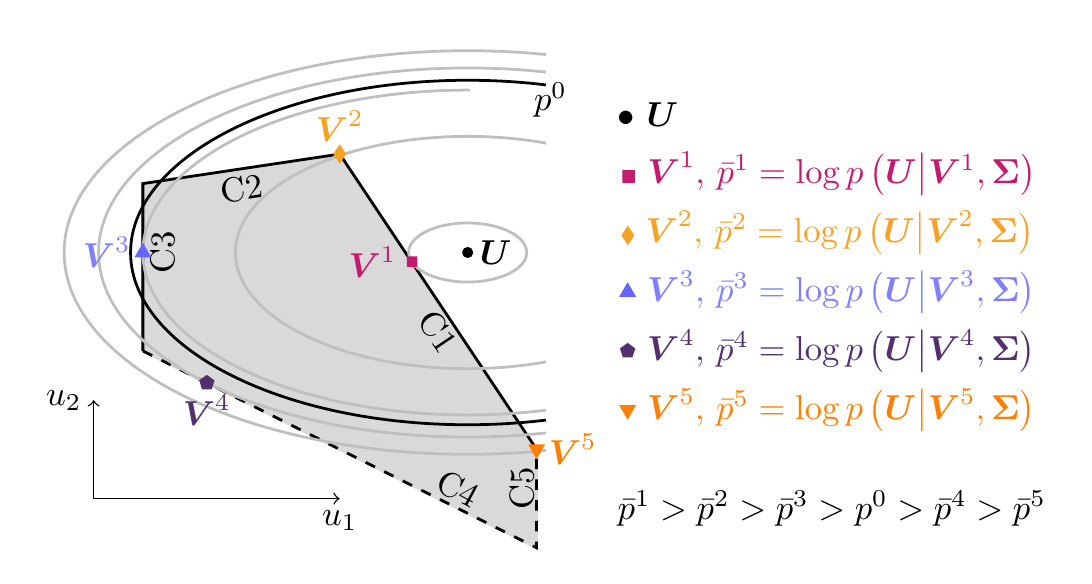}    
\caption{
Illustration of upper bounds on likelihoods $\bar p{}^i$ as computed in \eqref{ifac:eq:constraint_likelihood} for polytopic constraints $i=1,...5$.
The five upper bounds $\bar p{}^1$--$\bar p{}^5$ (gray ellipses) for the five constraints (C1--C5) as well as $p^0$ (black ellipse) are displayed as level sets.
For C$i$, $\V^i$ denotes the corresponding, projected input sequence.
} 
\label{ifac:fig:proj}
\end{center}
\end{figure}

Fig.~\ref{ifac:fig:proj} illustrates the concept of the upper bounds on the constraint likelihoods (Line~\ref{ifac:alg:barp} in Algorithm~\ref{ifac:alg:ML}).
In the given example, constraint 4 and constraint 5 do not have to be considered for learning, i.e., do not need to be enumerated. 
The resulting candidate constraint combinations are
\begin{align*}
\mathcal D
=
\left\{
\hspace{-0.15cm}
\left\{
\hspace{-0.1cm}
\begin{bmatrix}
1 \\ 0 \\ 0 \\ 0 \\ 0
\end{bmatrix},\tilde p{}^1
\right\},
\left\{
\hspace{-0.1cm}
\begin{bmatrix}
0 \\ 1 \\ 0\\ 0 \\ 0
\end{bmatrix},\tilde p{}^2
\right\},
\left\{
\hspace{-0.1cm}
\begin{bmatrix}
1 \\ 1 \\ 0\\ 0 \\ 0
\end{bmatrix},\tilde p{}^3
\right\},
\left\{
\hspace{-0.1cm}
\begin{bmatrix}
0 \\ 0 \\ 1\\ 0 \\ 0
\end{bmatrix},\tilde p{}^4
\right\}
\hspace{-0.15cm}
\right\}
\end{align*}
Note that $\tilde p{}^1=\bar p{}^1$, $\tilde p{}^2=\bar p{}^2$, $\tilde p{}^3=\bar p{}^2$, and $\tilde p{}^4=\bar p{}^3$.

%%%%%%%%%%%%%%%%%%%%%%%%%%%%%%%%%%%%%%%%%%%%%%%%%%%%%%%%%%%%%%%%%%%%%%%%%%%%%%%%%%%%%%%%%%%%%%%%%%%%%%%%%%%%%%%%%%%%%%%%%%%%%%%%%%%%%%%%%%%%%%%%%%%%%%%%%%%%%%%%%%%%%%%%%%%%%%%%%%%%%%%%%%%%%%%%%%%%%%%%%%%%%%%%%%%%%%%%%%%%%%%%%%%%%%%%%%%%%%%%%%%%%%%%%%%%%%%%%%%%%%%%%%%%%%%%%%%%%%%%%%%%%%%%%%%%%%%%%%%%%%%%%%%%%%%%%%%%%%%%%%%%%%%%%%%%%%%%%%%%%%%%%%%%%%%%%%%%%%%%%%%%%%%%%%%%%%%%%%%%%%%%%%%%%%%%%%%%%%%%%%%%%%%%%%%%%%%%%%%%%%%%%%%%%%%%%%%%%%%%%%%%%%%%%%%%%%%%%%%%%%%%%%%%%%%%%%%%%%%%%%%%%%%%%%%%%%%%%%%%%%%%%%%%%%%%%%%%%%%%%%%%%%%%%%%%%%%%%%%%%%%%%%%%%%%%%%%%%%%%%%%%%%%%%%%%%%%%%%%%%%%%%%%%%%%%%%%%%%%%%%%%%%%%%%%%%%%%%%%%%%%%%%%%%%%%%%%%%%%%%%%%%%%%%%%%%%%%%%%%%%%%%%
\section{Analysis for Linear Systems and Quadratic Cost Function}
In this section, we present properties of the learning methods for a common class of dynamical systems and cost functions.
Suppose the system in \eqref{ifac:eq:sys} is unconstrained and linear (time-invariant or time-varying), i.e., 
\begin{subequations}
\begin{equation}
\label{ifac:eq:sys_ex}
\f_k(\x_k,\boldsymbol u_k)=\boldsymbol A_k\x_k+\boldsymbol B_k\boldsymbol u_k,
\end{equation} 
and the cost function is of the form 
\begin{equation} 
\label{ifac:eq:cost_ex}
\pa^T \boldsymbol \phi(\boldsymbol V,\boldsymbol Z)=
\sum_{k=0}^N \boldsymbol z_k^T \boldsymbol Q_{\pa} \boldsymbol z_k + \boldsymbol v_k^T \boldsymbol R_{\pa} \boldsymbol v_k
\end{equation}
\end{subequations}
with $\boldsymbol Q_{\pa},\boldsymbol R_{\pa}\succ \0$.
Then, the stationarity condition can be written as 
\begin{align*}
\nabla_{\V}\mathcal L_{\pa}(\V,\z(0))=\M_{\pa}\V + \N_{\pa}\z(0),
\end{align*}
where both $\M_{\pa}$ and $\N_{\pa}$ depend on the system dynamics, i.e., $\boldsymbol A_k,\boldsymbol B_k$, and are linear in their parameters 
$\pa$, i.e., 
$\M_{\mu \pa_1+\pa_2}=\mu \M_{\pa_1}+\M_{\pa_2}$ and $\N_{\mu \pa_1+\pa_2}=\mu \N_{\pa_1}+\N_{\pa_2}$ with the scalar $\mu$. 

Let $\gt$ be the true parameters and, without loss of generality, $\|\gt\|_2=1$ (scale-invariance of the cost function), i.e., $\nabla_{\boldsymbol V}\mathcal L_{\gt}(\boldsymbol V,\boldsymbol z(0))=\boldsymbol M_{\gt}\boldsymbol V + \boldsymbol N_{\gt}\boldsymbol z(0)=\0$.
The goal of the learning methods is thus to estimate $\gt$ or any scaled version $\est =\mu \gt$ with $\mu > 0$ from data. 
Desirable properties of the learning method are that $\gt$ results in expectation and that all $\est =\mu \gt$ with $\mu > 0$ are equally likely.

Theorem~\ref{thm:variant2} shows that the expected value of Method~3 is the true parameter vector $\est =\gt$ and that Method~3 is indifferent toward the cost function's scale, i.e., any $\est =\mu \gt$ with $\mu > 0$ are equally likely (in expectation).
Method~2 is equally indifferent toward the parameters' scale but the expected parameters are only $\est =\gt$ if $\x(0)=\z(0)$ (proof omitted as it can be similarly derived).
Theorem~\ref{thm:variant0} shows that the expected parameters $\est$ of Method~1 are not necessarily $\gt$ and that Method~1 is not indifferent toward the parameters' scale.

\begin{thm}   
\label{thm:variant2}\quad
Consider unconstrained, linear systems of the form 
\eqref{ifac:eq:sys_ex}
and cost functions
\eqref{ifac:eq:cost_ex}.
Let $\boldsymbol U\sim \mathcal N(\boldsymbol V,\boldsymbol \Sigma)$ and $\boldsymbol x(0)\sim \mathcal N(\boldsymbol z(0),\boldsymbol \Sigma_0)$.
In expectation, Method~3 returns $\gt$ (result~1) and any other parameter realization is necessarily $\est \propto \gt$ (result~2).
\end{thm}

\textbf{Proof.} 
\quad 
Without loss of generality, define $\pa = \gt + \mu  \parpa$ with $\mu\in \mathbb{R}$ and $\parpa \in \mathbb{R}^p$ such that $\|\parpa\|_2=1$. 
The results will be shown by proving the following statements: 

\textit{Claim 1:} For any $\parpa$, 
\begin{align*}
\hat
\mu=0 = \arg \max_{\mu,\V}\
&
\mathbb{E}%_{\U,\x(0)}
\left[
-
\|
\U - \V
\|_{\boldsymbol \Sigma^{-1}}
-
\|
\x(0) - \z(0)
\|_{\boldsymbol \Sigma_0^{-1}}
\right]
\\
{\rm s.t.}\ & 
\0 =\M_{\gt + \mu\parpa}\V  + \N_{\gt + \mu\parpa} \z(0)
\end{align*}

\textit{Claim 2:} For $\parpa=\gt$, any $\mu\in \mathbb{R}$ minimizes
\begin{align*}
\hat \mu=
\arg \max_{\mu,\V}\
&
\mathbb{E}%_{\U}
\left[
-
\|
\boldsymbol U - \boldsymbol V
\|_{\boldsymbol \Sigma^{-1}}
-
\|
\boldsymbol x(0) - \boldsymbol z(0)
\|_{\boldsymbol \Sigma_0^{-1}}
\right]
\\
{\rm s.t.}\ & 
\0 =\M_{\gt + \mu\parpa}\V  + \N_{\gt + \mu\parpa} \z(0)
\end{align*}

Result~1 follows readily from Claim~1 as $\pa=\gt$.
Result~2 follows from Claim~2 as $\pa=(1+\hat \mu)\gt \propto \gt$.

\textbf{Proofs of Claim 1 and Claim 2.} \quad
Notice first that $\boldsymbol M_{\pa}\succ \boldsymbol 0$ is invertible.
The log-likelihood of Method~3 in \eqref{ifac:eq:inverse2}  is proportional to 
\begin{align}
\label{ifac:eq:proof_temp}
-\|
\U - \V
\|_{\boldsymbol \Sigma^{-1}}
-
\|
\boldsymbol x(0) - \boldsymbol z(0)
\|_{\boldsymbol \Sigma_0^{-1}} .
\end{align}
Using $\boldsymbol V=-\boldsymbol M_{\pa}^{-1}\boldsymbol N_{\pa} \boldsymbol z(0)$, \eqref{ifac:eq:proof_temp} can be written as
\begin{align}
\label{ifac:eq:proof1}
-
\|
\boldsymbol M_{\pa}\boldsymbol U + \boldsymbol N_{\pa} \boldsymbol z(0)
\|_{
\left(
\boldsymbol M_{\pa}^T
\boldsymbol \Sigma
\boldsymbol M_{\pa}
\right)^{-1}
}
-
\|
\boldsymbol x(0) - \boldsymbol z(0)
\|_{\boldsymbol \Sigma_0^{-1}}.
\end{align}
Then, using linearity 
($\M_{\pa}=\M_{\gt} + \mu \M_{\parpa}$ and
$\N_{\pa}=\N_{\gt} + \mu \N_{\parpa}$), 
$\U = \V +  \partial\V$, and $\x(0) = \z(0) +  \partial\boldsymbol z$,  the expected value of \eqref{ifac:eq:proof1} yields
\begin{equation}
\label{ifac:eq:proof2}
\begin{aligned}
-
\mu^2
\|
\boldsymbol M_{\parpa}\boldsymbol V + \N_{\parpa} \z(0)
\|_{
\left(
\boldsymbol M_{\pa}^T
\boldsymbol \Sigma
\boldsymbol M_{\pa}
\right)^{-1}
}
\\
-
{\rm trace}
\left(
\boldsymbol \Sigma \boldsymbol \Sigma^{-1}
\right)
-
{\rm trace}
\left(
\boldsymbol \Sigma_0 \boldsymbol \Sigma_0^{-1}
\right)
\end{aligned}
\end{equation}
Therefore, for any $\parpa$, $\mu=0$ maximizes \eqref{ifac:eq:proof2}, which proves Claim 1.
For $\parpa=\gt$, $\M_{\parpa}\V + \N_{\parpa} \z(0) = \0$ and $\mu\in \mathbb{R}$ minimizes \eqref{ifac:eq:proof2}, which proves Claim 2.
\qed

\begin{thm}  
\label{thm:variant0}\quad
Consider unconstrained linear systems of the form \eqref{ifac:eq:sys_ex} and cost functions \eqref{ifac:eq:cost_ex}.
Let $\U\sim \mathcal N(\V,\boldsymbol \Sigma)$ and $\x(0)\sim \mathcal N(\z(0),\boldsymbol \Sigma_0)$.
Then, $\gt$ does not result in expectation from Method~1 (result~1).
Further, Method~1 is not scale-invariant (result~2).
\end{thm}

\begin{pf}   \quad
For the considered class of dynamical systems, the cost function in \eqref{ifac:eq:naive} is
\begin{align}
\label{ifac:eq:proof3}
\|
\M_{\pa} \U
+
\N_{\pa} \x(0)
\|_{\boldsymbol I}.
\end{align}
We define $\pa = \gt + \mu  \parpa$ with $\mu\in \mathbb{R}$ and $\parpa\in \mathbb{R}^p$ and $\|\parpa\|_2=1$.
Then, using linearity 
($\boldsymbol M_{\pa}=\boldsymbol M_{\gt} + \mu \boldsymbol M_{\parpa}$ and
$\boldsymbol N_{\pa}=\boldsymbol N_{\gt} + \mu \boldsymbol N_{\parpa}$), 
$\boldsymbol U = \boldsymbol V +  \partial\boldsymbol V$, and $\boldsymbol x(0) = \z(0) +  \partial\z$,  the expected value of \eqref{ifac:eq:proof3} yields
\begin{equation}
\label{ifac:eq:proof4}
\begin{aligned}
&\mathbb{E}
\left[
\|
\boldsymbol M_{\pa} \boldsymbol U
+
\boldsymbol N_{\pa} \boldsymbol x(0)
\|_{\boldsymbol I}
\right]
=
\\
& 
\mu^2
\left(
\|
\M_{\parpa}\V + \N_{\parpa} \z(0)
\|_{
\boldsymbol I
}
+ 
{\rm t}_{\parpa, \parpa}
\right)
+2\mu\
 {\rm t}_{\gt, \parpa}
+ {\rm t}_{\gt, \gt}
\end{aligned}
\end{equation}
with
\begin{align*}
{\rm t}_{\gt, \gt}
&=
{\rm trace}
(
\boldsymbol M^T_{ \gt} \boldsymbol M_{ \gt}\boldsymbol \Sigma
)
+ {\rm trace}
(
\boldsymbol N^T_{ \gt} \boldsymbol N_{ \gt}\boldsymbol \Sigma_0
)
\\
{\rm t}_{\gt,\parpa}
&=
{\rm trace}
(
\boldsymbol M^T_{ \gt} \boldsymbol M_{\parpa}\boldsymbol \Sigma
)
+ {\rm trace}
(
\boldsymbol N^T_{ \gt} \boldsymbol N_{\parpa}\boldsymbol \Sigma_0
)
\\
{\rm t}_{\parpa,\parpa}
&=
{\rm trace}
(
\boldsymbol M^T_{\parpa} \boldsymbol M_{\parpa}\boldsymbol \Sigma
)
+ {\rm trace}
(
\boldsymbol N^T_{\parpa} \boldsymbol N_{\parpa}\boldsymbol \Sigma_0
).
\end{align*}
The optimal $\mu$  minimizing \eqref{ifac:eq:proof4} is not necessarily $0$ but a function of $\parpa$, which proves result~1:
\begin{align*}
\mu =
-
\frac{
{\rm t}_{\gt,\parpa}
}{\|
\boldsymbol M_{\parpa}\boldsymbol V + \boldsymbol N_{\parpa} \z(0)
\|_{
\boldsymbol I
}
+ 
{\rm t}_{\parpa,\parpa}
}.
\end{align*}
For $\parpa = \gt$, $\mu=-1$ and $\pa = \gt -\gt=\0$, i.e., the estimator is not scale-invariant, which proves result~2.
\qed
\end{pf}

\section{Simulation Results}
\label{ifac:sec:results}
In this section, we utilize the three discussed methods for learning the cost function's parameters.

\subsection{Simulation Setup}
\subsubsection{System dynamics and constraints}
We consider one linear system and one nonlinear system with dynamics
\begin{subequations}
\begin{align}
\label{ifac:eq:sysL}
\boldsymbol x(k+1) 
&=
\begin{bmatrix}
1 
&
1
\\
0
& 
1
\end{bmatrix}
\boldsymbol x(k)
+
\begin{bmatrix}
0
\\
1
\end{bmatrix}
 u(k)
\\
\label{ifac:eq:sysNL}
\boldsymbol x(k+1) 
&=
\begin{bmatrix}
1 
&
1 - u(k)^2
\\
0
& 
1
\end{bmatrix}
\boldsymbol x(k)
+
\begin{bmatrix}
0
\\
1
\end{bmatrix}
u(k).
\end{align}
\end{subequations}
The inputs are constrained as
$|u(k)|\leq 1$.

\subsubsection{Cost function}
The true cost function is chosen as 
\begin{align*}
\gt^T \boldsymbol \phi(\boldsymbol V,\boldsymbol Z)
\end{align*}
with 
\begin{align*}
\gt= 
\begin{bmatrix}
1 \\ 1 \\ 1 \\ 1
\end{bmatrix},\quad
\boldsymbol \phi(\V,\boldsymbol Z)
=
\begin{bmatrix}
\textstyle 
\sum_{k=0}^{N}
z_{1,k}^2
\\
\textstyle 
\sum_{k=0}^{N}
z_{2,k}^2
\\
\textstyle 
\sum_{k=0}^{N-1}
(v_{k+1}-v_k)^2
\\
\textstyle 
\sum_{k=0}^{N}
v_k^2
%V^T V
\end{bmatrix}
%,  \z_k=\begin{bmatrix} z_{1,k}\\ z_{2,k} \end{bmatrix}
\end{align*}
with $\z_k=[z_{1,k}\ z_{2,k}]^T$ 
and $N=10$.
As the cost function is scale-invariant, we fix one parameter and learn 
\begin{align*}
\est{}^T= 
\begin{bmatrix}
\hat \theta_1 & \hat \theta_2 & \hat \theta_3 & 1
\end{bmatrix}.
\end{align*}

\subsubsection{Data generation}
In order to generate the data, we sample the initial conditions $\z(0)$ from $\z(0) \sim \mathcal{N}(\0,\boldsymbol I)$.
Using $\z(0)$, we obtain the optimal input sequence $\V$ using \eqref{ifac:eq:forward}.
Then, the (sub-optimal) demonstration is generated as $\U\sim \mathcal{N}(\V, \sigma_u^2 \boldsymbol I_{10})$ and $\x(0)\sim \mathcal{N}(\boldsymbol z(0),\sigma_0^2 \boldsymbol I_{2})$, where $\sigma_u$ and $\sigma_0$ are varied logarithmically as
\begin{align*}
\sigma_u,\sigma_0
\in
\{
0.0001 
,
0.000215
,
0.000464
,
0.001 
,\
\\
0.00215
,
0.00464
,
0.01 
,\
\\
0.0215
,
0.0464
,
0.1 
,\
\\
0.215
,
0.464
,
1
\}
\end{align*}

\subsubsection{Evaluation criterion}
We evaluate the learned parameters by comparing $\V$ resulting from \eqref{ifac:eq:forward} with $\gt$ and $\hat{\boldsymbol{V}}$ resulting from \eqref{ifac:eq:forward} with $\est$ as
\begin{align}
\label{ifac:eq:error}
{\rm error}
=
\frac{
\|
\hat{\boldsymbol{V}} - \boldsymbol{V}
\|_2
}{
\|
\boldsymbol{V}
\|_2}.
\end{align}

\subsection{Learning Results}
For every tuple $\{\sigma_u,\sigma_0\}$, we repeat the data generation process 1000 times and learn $\hat{\boldsymbol \theta}$ using the three inverse learning methods.

\begin{figure}[t!]
\begin{center}
\includegraphics[trim={0 0 0 0},clip,width=0.975\columnwidth]{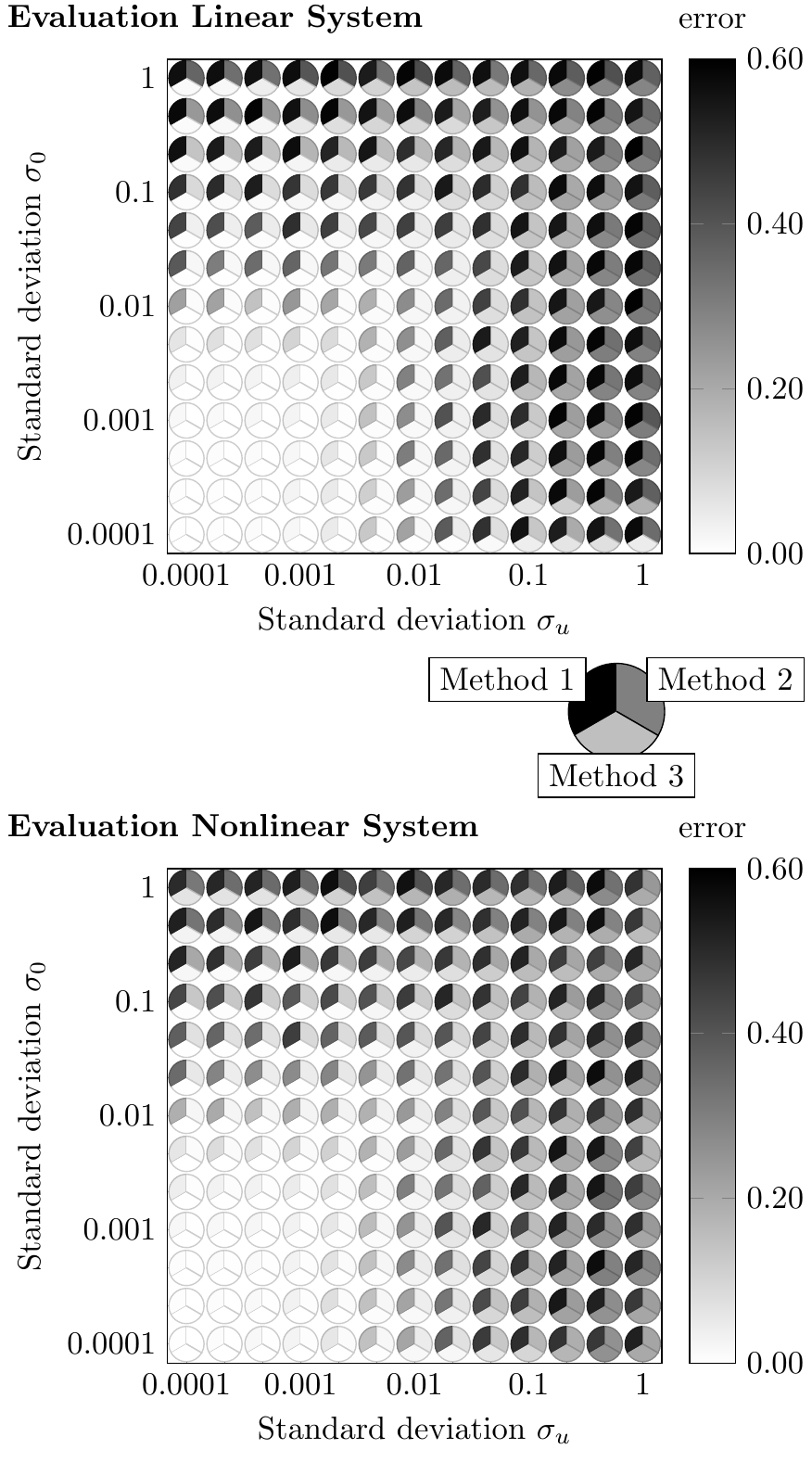}    
\caption{
Median of ${\rm error}\in [0,0.6]$ for different standard deviations $\sigma_u$ and $\sigma_0$ (color map from white to black) for the three inverse learning methods.
} 
\label{ifac:fig:heat}
\end{center}
\end{figure}

Fig.~\ref{ifac:fig:heat} illustrates the median of the error in \eqref{ifac:eq:error} for the 1000 trials for $\{\sigma_u,\sigma_0\}$ and the three learning methods.
For $\sigma_u=\sigma_0=0$, ${\rm error}=0$ for all three methods.
In the presence of noise $\sigma_u,\sigma_0>0$, the learning results degrade differently.
The error increases for larger standard deviations $\{\sigma_u,\sigma_0\}$ for all three methods.
However, it can be seen that with increased noise levels (sub-optimal data), the error increases more quickly for Method~1, whereas the errors remain smaller for Method~2 and Method~3.
Now, consider small $\sigma_u$. 
For increased $\sigma_0$, the error for Method~3 is significantly smaller than Method~2, which is expected since Method~3 optimizes over $\z(0)$. 
For small $\sigma_0$ and increasing $\sigma_u$, Method~3 also outperforms Method~2, which suggests that optimizing over the initial condition is advantageous even for small uncertainties in the initial conditions.
Note that the standard deviations $\sigma_u=\sigma_0>0.1$ are unrealistically high as $|u|\leq 1$.

\begin{figure*}[h]
\begin{center}
\includegraphics[trim={0 0.25cm 0.9cm 0},clip,width=1.7\columnwidth]{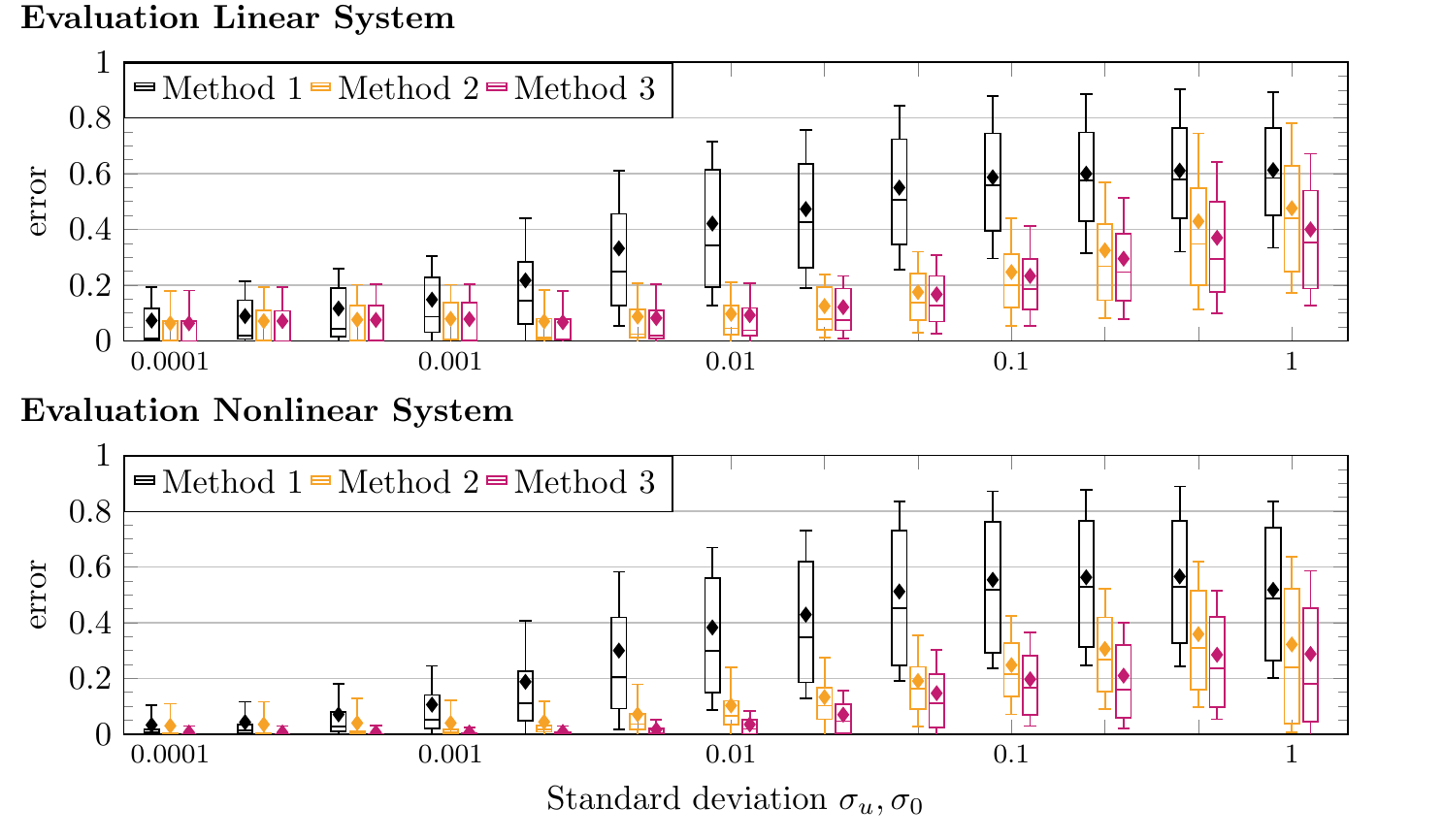}    
\caption{Evaluation of the error in \eqref{ifac:eq:error} for the three inverse learning methods for different noise levels $\sigma_u=\sigma_0$. 
The diamond symbol and vertical line represent the mean and the median, respectively; the box edges represent the 25th and the 75th percentiles; and the whiskers represent the standard deviation.
} 
\label{ifac:fig:stat}
\end{center}
\end{figure*}
Fig.~\ref{ifac:fig:stat} shows a more detailed statistical evaluation of the error in \eqref{ifac:eq:error} for the 1000 trials and $\sigma_u=\sigma_0$. 
First, it can be seen that the estimation with Method~2 and Method~3 have a lower error compared to Method~1.
The errors tend to be lower for the nonlinear system compared to the linear system, which can best be seen for $\sigma_0=\sigma_u<0.01$.
The relatively low errors of Method~2 and Method~3 suggest the superiority of the maximum likelihood formulation over the convex relaxation approach of Method~1 measured with respect to the predictive performance, i.e., \eqref{ifac:eq:error}.

\subsection{Computation Time}
Table~\ref{ifac:tb:comp} states the median computation time for all samples of the initial conditions for the three learning methods using MATLAB with 
the hardware configuration: 3.1~GHz Intel Core i7, 16 GB
1867 MHz DDR3, and Intel Iris Graphics 6100 1536 MB.
Method~1 is convex and, therefore, the computation is cheap and requires less than one second.
Method~2 is computationally slightly more involved but can still be solved in around one second.
Method~3 is more demanding as also the initial condition, $\z(0)$, is an optimization variable.

\begin{table}[!h]
\begin{center}
\caption{Computation time}\label{ifac:tb:comp}
\begin{tabular}{llll}
System &  Method & Median over all samples  \\\hline
\multirow{3}{*}{\rotatebox{0}{Linear \eqref{ifac:eq:sysL}}}
& Method~1 & $T_{\rm L,M1}=0.148$s   \\
& Method~2 & $T_{\rm L,M2}=1.19$s  ($8.04T_{\rm L,M1}$) \\
& Method~3 & $T_{\rm L,M3}=3.07$s  ($20.8T_{\rm L,M1}$) \\
\\
\vspace{-0.6cm}
\\
\multirow{3}{*}{\rotatebox{0}{Nonlinear \eqref{ifac:eq:sysNL}}}
& Method~1 & $T_{\rm NL,M1}=0.142$s   \\
& Method~2 & $T_{\rm NL,M2}=0.584$s  ($4.12T_{\rm NL,M1}$) \\
& Method~3 & $T_{\rm NL,M3}=1.44$s  ($10.2T_{\rm NL,M1}$) \\ \hline
\end{tabular}
\end{center}
\end{table}

\begin{figure*}[h]
\begin{center}
\includegraphics[trim={0 0.25cm 0.9cm 0},clip,width=1.77\columnwidth]{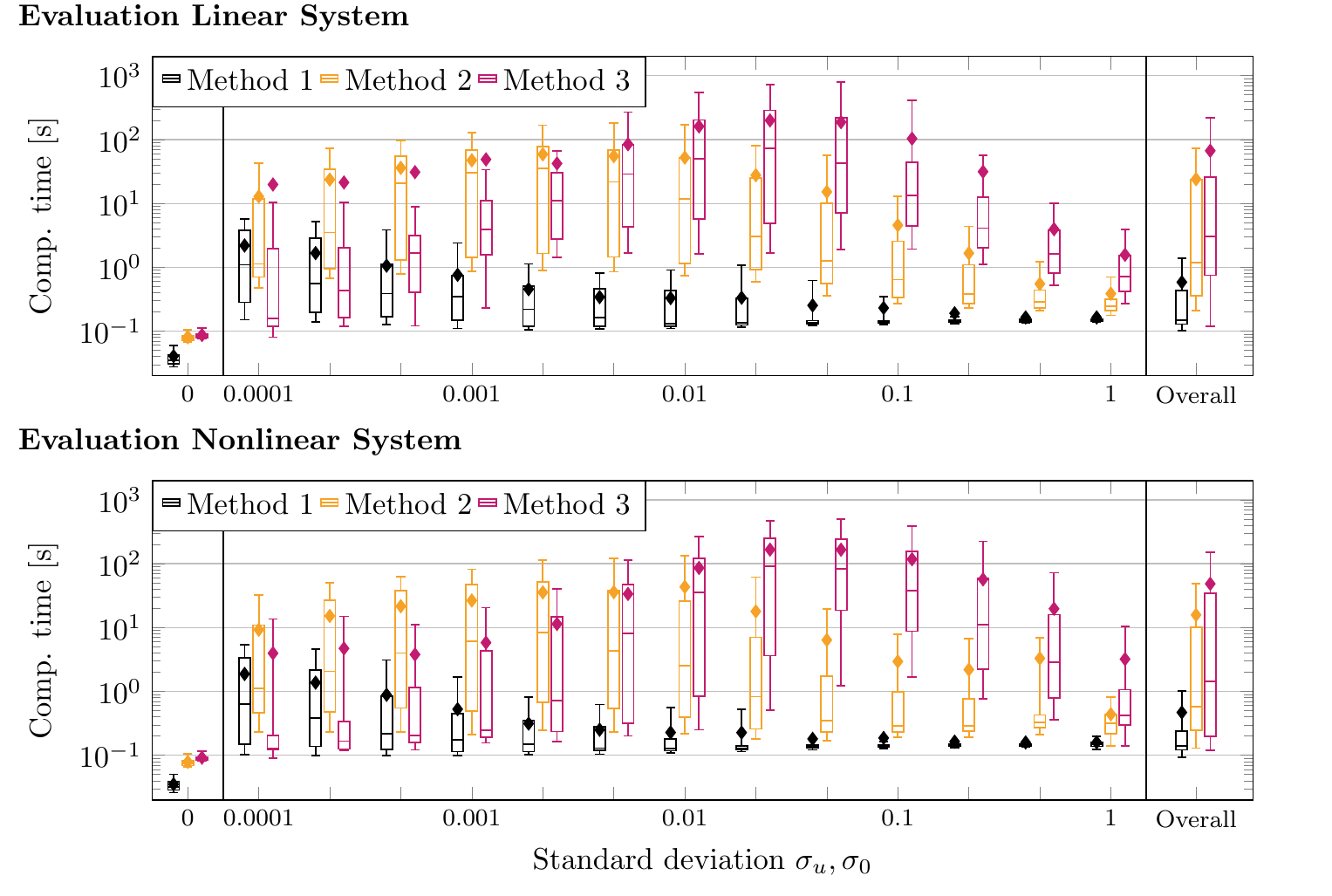}    
\caption{Statistical evaluation of the computation time for the three inverse learning methods with $\sigma_u=\sigma_0$.
The diamond symbol and vertical line represent the mean and the median, respectively; the box edges represent the 75th and the 25th percentiles; and the whiskers represent the 90th and the 10th percentiles.
} 
\label{ifac:fig:comp}
\end{center}
\end{figure*}

Fig.~\ref{ifac:fig:comp} shows a more detailed statistical evaluation of the computation times for $\sigma_u=\sigma_0$.
The  three learning methods have their respective peak computation times at different noise levels, i.e., Method~1's maximum computation times are highest for lower noise levels (peak of median at $\sigma_u=\sigma_0=0.0001$); Method 2's maximum times occur for slightly higher noise levels (peak at $\sigma_u=\sigma_0=0.00215$); whereas Method~3's peak is for high noise levels (peak at $\sigma_u=\sigma_0=0.0215$). 
For all methods and noise levels, the mean value is higher than the median, which is expected as the median is less susceptible to outliers, i.e., instances with particularly long computation times.

\section{Conclusion}
This paper presented three inverse optimal control methods; 
one method that uses a convex relaxation of the KKT optimality conditions and two methods that combine the KKT conditions with maximum likelihood estimation.
It proposed a branch-and-bound-style algorithm for the maximum likelihood formulation, which is based on likelihood arguments to systematically deal with constraints in the presence of noisy data.
A simulation study exemplified the three inverse learning methods with both a constrained, linear and a nonlinear system.
The results showed that the likelihood estimation methods can be implemented quite efficiently and yield robust learning results, whereas the convex method is computationally efficient but less robust to noise in the training data.

\addtolength{\textheight}{-2.5cm}

%\bibliography{IEEEabrv,menner}         
\bibliography{ifac.bbl}

\end{document}